\documentclass[aps,prb,preprint,showpacs,superscriptaddress]{revtex4-1}

\usepackage{graphicx,subfigure}
\usepackage{dcolumn}
\usepackage{bm}
\usepackage{ctable} 

\begin{document}


\title{Structure and dynamics of the fullerene polymer Li$_4$C$_{60}$ \\
studied with neutron scattering}


\author{S. Rols}\email[Corresponding author, ]{rols@ill.eu}
\affiliation{Institut Laue Langevin, BP 156, 6 rue Jules Horowitz, 38042 Grenoble Cedex, France}

\author{D. Pontiroli}\email[Corresponding author, ]{Daniele.pontiroli@fis.unipr.it}
\homepage{http://www.fis.unipr.it/daniele.pontiroli/}
\affiliation{Dipartimento di Fisica, Universit\`a di Parma, Via G. Usberti 7/a, 43100 Parma, Italy}

\author{C. Cavallari}
\affiliation{Institut Laue Langevin, BP 156, 6 rue Jules Horowitz, 38042 Grenoble Cedex, France}
\altaffiliation[also at: ]{Dipartimento di Fisica e Scienze della Terra, Universit\`a di Parma, Via G. Usberti 7/a, 43100 Parma, Italy}
\author{M. Gaboardi}

\author{M. Aramini}
\affiliation{Dipartimento di Fisica, Universit\`a di Parma, Via G. Usberti 7/a, 43100 Parma, Italy}

\author{D. Richard}
\author{M. R. Johnson}
\affiliation{Institut Laue Langevin, BP 156, 6 rue Jules Horowitz, 38042 Grenoble Cedex, France}
\author{J. M. Zanotti}
\affiliation{Laboratoire L\'{e}on Brillouin, CEA Saclay, 91191 Gif Sur Yvette Grenoble Cedex, France}
\author{E. Suard}

\author{M. Maccarini}
\altaffiliation[Present adress: ]{Team SyNaBi, Laboratory TIMC/IMAG, University Grenoble Alpes, UMR CNRS 5525, Pavillon Taillefer, 38700 La Tronche, France}
\affiliation{Institut Laue Langevin, BP 156, 6 rue Jules Horowitz, 38042 Grenoble Cedex, France}
\author{M. Ricc\'o}
\affiliation{Dipartimento di Fisica, Universit\`a di Parma, Via G. Usberti 7/a, 43100 Parma, Italy}

\date{\today}%

\begin{abstract}
The two-dimensional polymer structure and lattice dynamics of the superionic conductor Li$_4$C$_{60}$ are investigated by neutron diffraction and spectroscopy. The peculiar bonding architecture of this compound is confirmed through the precise localisation of the carbon atoms involved in the intermolecular bonds. The spectral features of this phase are revealed through a combination of \textit{ab-initio} lattice dynamics calculations and inelastic neutron scattering experiments. The neutron scattering observables are found to be in very good agreement with the simulations which predict a partial charge transfer from the Li atoms to the C$_{60}$ cage. The absence of a well defined band associated with the Li atoms in the experimental spectrum suggests that this species is not ordered even at the lowest temperatures. The calculations predict an unstable Li sublattice at a temperature of $\sim$ 200 K, that we relate to the large ionic diffusivity of this system: low frequency optic modes of the Li ions couple to the soft structure of the polymer.
\end{abstract}




\maketitle

\section{Introduction}
\label{sec:intro}
Solid fullerene polymerization is a well--established phenomenon \cite{Fischer}. C$_{60}$ can polymerize by undergoing high temperature and high pressure treatments, after ion irradiation or by exposure to light \cite{regueiro95}. In all these cases, fullerene units invariably connect together by four-membered carbon rings, via [2+2] cyclo--addition reactions \cite{Rao}. 

The intercalation of light electron donors (\textit{i.e.} alkali atoms Li or Na) in the voids of the host fullerene lattice can sometimes induce the formation of polymerized structures. In this case, the charge transfer from the metal to the highly electronegative C$_{60}$ provides the necessary \textit{chemical pressure} for adjacent buckyballs to be close enough together to establish intermolecular bonds. 
A large variety of 1D \cite{winter98, bendele98} and 2D \cite{oszlany97} fulleride polymer structures  was found when intercalated with small-alkali atoms. In addition to the aforementioned cycloaddition mechanism, fullerenes can connect through single C-C bonds. In the latter category, lithium intercalated fullerides Li$_{\mathrm{x}}$C$_{60}$ $(\mathrm{x} \leq 6)$, are remarkable because their structure consists of 2D polymeric fullerenes interconnected by a sequence of single C-C bonds and four-membered carbon rings, propagating along two orthogonal directions in the polymer plane \cite{ricco05}. Li$_4$C$_{60}$ belongs to this class of compounds, together with the isostructural alkali-earth intercalated Mg$_2$C$_{60}$ polymer, suggesting that the on-ball charge transfer is the key factor driving the polymerization arrangement in these systems \cite{pontiroli12}. The coexistence of the two bonding schemes was also observed in the \textit{fullerenium salt} C$_{60}$(AsF$_{6}$)$_{2}$. In this solid, the fullerene units are oxidized to the state C$_{60}^{2+}$, a rather unstable condition, which results in an unusual 1D zigzag polymerization \cite{ricco10}.\\
In addition to its novel structure, Li$_4$C$_{60}$ shows a very large ionic conductivity at low temperature, which is an exceptional property for a solid material. The ionic conductivity reaches the value of $\sigma \sim$ 10$^{-2}$ S/cm at room temperature, which is comparable to that observed in liquid electrolytes, indicating that this material could possibly find applications, for example, in fuel cell devices \cite{ricco09}. This physical property originates from the presence of intrinsic unoccupied sites interconnected by $3D$ pathways in the crystalline structure, allowing the diffusion of the Li$^+$ ions. The large amplitude movements of the $C_{60}$ cages \textit{e.g.} their rotations or radial deformations, can facilitate the Li$^+$ ions jumping from one site to another. Such dynamical disorder is intrinsically involved in the atomic mechanisms leading to oxygen diffusion in certain oxides, like the onset of Mo0$_x$ free rotations in Bi$_{26}$Mo$_{10}$O$_{69}$ \cite{Ling12}. In general the superionic character of a material follows an order-disorder transition which unlocks large amplitude movements. In Li$_4$C$_{60}$, no such transition is observed and the superionic nature of the material appears progressively in an apparently very ordered fullerene host structure.\\

It is possible to break the intermolecular bonds and recover the monomer phase through a moderate thermal annealing. The resulting phase shows an unexpected metallic behaviour \cite{ricco07} which appears to be in contrast with accepted theories indicating the A$_4$C$_{60}$ compounds (A = alkali metal) to be Mott-Jahn Teller insulators \cite{capone00, han00}.\\

In the present work we report on the use of neutron scattering to study Li$_4$C$_{60}$. The structure is investigated by neutron powder diffraction performed at low temperature, \textit{i.e}. in a regime where the Li diffusion is hindered. Rietveld refinement is used to precisely localise the carbon atoms of the distorted fullerene cage --especially those involved in the intermolecular bonds-- and to reveal the Li positions in the host lattice.
The structural information is then used to calculate the lattice dynamics of the polymeric system in the harmonic approximation. The phonon dispersion curves, the vibrational density of states and the thermodynamic properties of this molecular system are discussed. The validity of the model is illustrated by the excellent agreement between the neutron scattering observables derived from simulations and those observed in our inelastic neutron scattering experiments (INS).\\

Finally, we present the INS results obtained on this system in the monomeric phase at high temperature. The experimental data are interpreted with the help of DFT Molecular Dynamics simulations (MD-DFT).  

\section{Experimental Methods}
\label{sec:experiment}
Li$_4$C$_{60}$ samples were prepared  by thermal decomposition of 99\% isotopically pure $^{7}$Li azide. Details of the procedure are described elsewhere \cite{ricco05}. Isotopic enrichment was required to avoid neutron absorption  arising from the natural fraction of $^{6}$Li. Sample handling was performed in a controlled atmosphere (high vacuum or Ar glove box with oxygen and moisture values below 1 ppm). 

The neutron powder diffraction measurements were performed at the Institut Laue Langevin (ILL), using the Super D2B two-axis diffractometer, operating in Debye-Scherrer geometry and in high resolution mode (neutron wavelength $\lambda$ = 1.59432 \AA). A 1.5 gr sample was sealed in a cylindrical vanadium can with indium o-ring. Measurements were performed using a Closed Cycle Refrigerator allowing data to be collected at 3.5 K, with an accumulation time of 10 hours.

Inelastic neutron scattering experiments were performed at the ILL, on the thermal time of flight spectrometer IN4C, and on the Filter Analyser Spectrometer IN1BeF. Additional measurements were performed at the Laboratoire L\'{e}on Brillouin (CEA Saclay) on the cold TOF spectrometer Mibemol, to follow the evolution of the dynamics during the depolymerisation process at high temperature. On IN4C, several incident wavelengths were used (1.1, 1.4, 1.5 and 2.4 \AA), allowing the dynamics to be measured either in Stokes (phonon creation) or in anti-Stokes (phonon annhilation) regimes: using short wavelengths, the dynamics can be measured at low temperatures in an extended energy range,  while using a relatively long wavelength, the dynamics can be measured in an extended energy range in the anti-Stokes regime, albeit at relatively high temperature. On IN1BeF, the measurements were conducted at low T (10 K). In both experiments, the 1.5 g sample was sealed inside an Al flat container with an Indium seal. On Mibemol the 1.5 gr sample was sealed inside a cylindrical niobium can and heated up to 800 K during 12 hours using a high temperature furnace. The incident neutron wavelength was 5 \AA\ giving the maximum neutron flux. The spectra were recorded as a function of time in order to follow the de--polymerisation process.

\section{Simulations}
\label{sec:simulations}
Total energy calculations, geometry relaxations and molecular dynamics simulations were performed using the projector--augmented wave (PAW) formalism \cite{paw_1,paw_2} of the Kohn-Sham density functional theory \cite{dft_1,dft_2}, within the generalized gradient approximation (GGA), implemented in the Vienna \textit{ab initio} simulation package (VASP) \cite{vasp_1, vasp_2}. The GGA was formulated by the Perdew--Burke--Ernzerhof (PBE) density functional \cite{pbe_1,pbe_2}. For total energy calculations, the electronic calculations were performed at the gamma point (k=0) and an energy cut--off of 499 eV was used. For the geometry relaxations, the energy cut--off was increased by 30\% and a 2x2x1 Monkhorst-Pack \cite{mp} $\vec k$ grid scheme was used for increased precision. The break conditions for the self-consistent field (SCF) and for the ionic relaxation loops were set to 10$^{-5}$ eV and 10$^{-6}$ eV \AA$^{-1}$, respectively. The latter break condition means that the obtained Hellmann--Feynman forces are less than 10$^{-6}$ eV \AA$^{-1}$ for the final geometries (futher details about the way such calculations are performed can be found in Ref.~\onlinecite{pald_2}).
Both full (lattice constants and atomic positions) and partial (only atomic positions) geometry relaxations were performed. Total energies were calculated for the 108 structures resulting from the individual displacements (atomic displacement = 0.03 \AA) of the 18 symmetry inequivalent atoms in the monoclinic (I2/m) phase, along the three inequivalent Cartesian directions ($\pm$x, $\pm$y and $\pm$z). Phonons are extracted from these calculations using the direct method  as implemented in the Phonon software \cite{phonon_1,phonon_2}. The phonon spectra obtained using the fully and partially relaxed structures differ only slightly, so that only the results reported in this paper concern the fully relaxed structure. \\
The charge distribution was calculated from the VASP relaxed geometries using the Bader scheme analysis \cite{bader}. We used the code \textit{bader} developed by Henkelmann's group \cite{tang09} which gives the partial charge on each atom.\\ 
The results of the lattice dynamics calculations will be discussed based on the usual properties: phonon dispersion curves, total and partial phonon density of states. In particular, a weighted phonon dispersion curve image is also discussed: for each phonon $|j\vec{q}\rangle$ characterised by energy $\hbar\omega_j(\vec{q})$ and momentum $\hbar\vec{q}$, an intensity is calculated as:

\begin{equation}
I_{\{\mu\}}(j,\vec{q})=\sum_{\{\mu\}}\|\vec{e_j}(\vec{q}\mid \mu)\|^2
\end{equation}
with $\vec{e_j}(\vec{q}\mid \mu)$ being the polarisation vector of the phonon $|j\vec{q}\rangle$ for atom $\mu$ in the unit cell. This identifies the atoms involved in each mode, and we will call this quantity $I_{\{\mu\}}(j,\vec{q})$ the \textit{participation factor} of the ensemble of atoms $\{\mu\}$ in the phonon mode $|j\vec{q}\rangle$.
  
Several thermodynamic functions (heat capacity, entropy, free energy and mean square atomic displacement) obtained in the framework of the harmonic approximation \cite{amermin} are also presented and discussed. These quantities are extracted from the calculations using the Phonon Software \cite{phonon_1}.\\
The neutron observable $S(Q, \omega)$ and $G(\omega)$ are calculated for a powder according to the PALD method \cite{pald,pald_2}.
Based on a set of modes $|j\vec{q}\rangle$, the coherent dynamical structure factor in the "one--phonon" approximation \cite{squires,mlovesey} $S_{coh}(\vec{Q}, \omega)$ is computed for a large number of $\vec{Q}=\vec{q}+\vec{G}_{hkl}$, with random orientation on a dense equidistant $\Vert \vec{Q} \Vert$ grid. $\vec{G}_{hkl}$ represents the vectors of the reciprocal lattice of the crystal. In phonon creation (Stokes):
\begin{equation}
S_{coh}(\vec{Q}, \omega)\propto \frac{n(\omega,T)+1}{\omega}\sum_{j}F_{j}(\vec{Q})\delta(\omega-\omega_j(\vec{Q}))
\end{equation}
with the one phonon form factor:
\begin{equation}
F_{j}(\vec{Q})= \Vert \sum_{d=1}^{N_d} \frac{b_d^{coh}e^{-W_d(\vec{Q})}}{\sqrt{M_d}}\vec{Q}\cdot \vec{e_j}(\vec{Q}\mid d)) \Vert ^2
\end{equation} 
and the Bose thermal population factor $n(\omega,T)=(exp(\hbar\omega/k_\beta T)-1)^{-1}$ with $k_\beta$ the Boltzmann constant. Index $d$ refers to the $d$th atom among the $N_d$ atoms in the unit cell, with mass $M_d$ and coherent scattering length $b_d^{coh}$. $W_d(\vec{Q})$ is the Debye-Waller factor of atom $d$. In a final step, $S_{coh}(\vec{Q}, \omega)$ is regrouped according to an adjustable $\Delta\Vert \vec{Q}\Vert$ and $\Delta\omega$ and orientationally averaged to obtain the isotropic function $S_{coh}(Q, \omega)$. \\
If the incoherent scattering length of carbon is negligible, that of lithium is not and one has to account for this scattering in the calculations. The Li total scattering cross section is therefore expressed as $S(Q, \omega)=S_{coh}(Q, \omega) + S_{inc}(Q, \omega) $ with the powder--averaged, one--phonon, incoherent cross section approximated to:
\begin{equation}
S_{inc}(Q, \omega) \propto N_dQ^2\frac{n(\omega,T)+1}{\omega}\sum_{d=1}^{N_d} \frac{(b_d^{inc})^2 e^{-2W_d(Q)}}{M_d}g_d(\omega)
\end{equation}
with $g_d(\omega)$ being the partial phonon density of state of atom $d$:
\begin{equation}
g_d(\omega)=\frac{1}{3NN_d}\sum_{j\vec{q}}\Vert \vec{e_j}(\vec{q}\mid d)) \Vert ^2 \delta(\omega-\omega_j(\vec{q}))
\end{equation}
and $N$ is the number of unit cells in the crystal (\textit{i.e.} \ $3NN_d$ is the total number of degrees of freedom in the sample).
The calculated $S(Q, \omega)$ spectra were subsequently adapted to the experimental conditions (temperature, instrument dependent scattering $(Q, \omega)$ range, resolution function) to allow for a direct comparison with the data. In particular, the INS-weighted spectra were convoluted with a Gaussian with energy-dependent width.
The calculated and experimental $S(Q, \omega)$ spectra were therefore treated the same way. In particular, the grouping and treatment performed in order to derive the generalised density of states (GDOS) is identical for both sets of data.\\

The structure of the high temperature monomeric phase of Li$_4$C$_{60}$ was taken from Ref.\onlinecite{ricco07}. As this structure is orientationally disordered, lattice dynamics simulations cannot be performed. We used DFT molecular dynamics simulations to investigate the atomic displacements in this phase. The MD simulations were performed at 800 K using a Nose-Hoover thermostat \cite{hoover85} as implemented in the VASP package. The MD time step was set to 1 fs and the total simulation time was 3 ps. The neutron weighted density of states were derived from the trajectories as the Fourier transform of the atomic velocity auto--correlation function. This analysis was performed using the nMoldyn package \cite{nmoldyn}. The simulation box used is restricted to the I$_{4/mmm}$ unit cell, \textit{i.e.} containing only 2 molecular units. As periodic boundary conditions were applied, artificial correlations between the rotating C$_{60}$ units can arise. However these mostly affect the quasielastic scattering region of the neutron spectra, which we do not discuss in the present paper.

\section{Results and discussion}
\label{sec:results}

\subsection{Neutron diffraction}
\label{ssec:Diffr}
Diffraction data were collected within the angular range 2$\theta$ = $10^{\circ}$ - $145^{\circ}$ with a step of $0.05^{\circ}$ (high resolution mode). 
Rietveld analysis was performed using the GSAS+EXPGUI software. The starting peak profile was determined by a preliminary Le Bail pattern decomposition and the background contribution was modelled with a polynomial curve containing 21 coefficients.
The structural analysis was started without considering the Li ions, due to the low (negative) scattering length of $^{7}$Li with respect to C. The initial fullerene arrangement was obtained by considering the carbon positions of the polymeric chains in KC$_{60}$ (where C$_{60}$ are connected by [2+2]-cycloaddition bonds), as described elsewhere \cite{guerrero98}.
Then, the fullerene chains were correctly rotated around their axis and the necessary transformations were applied to the atomic coordinates, in order to match the correct symmetry and the cell dimensions of Li$_4$C$_{60}$ ($a$ = 9.329(1) \AA, $b$ = 9.054(1) \AA, $c$ = 14.984(1)\AA, $\beta$ = 90.91(1)$^{\circ}$, space group I2/m), with the [2+2] polymerisation running along the $b$-axis \cite{margadonna04}.
At this stage, we found that the minimum in the $R_\mathrm{wp}$ agreement factor was reached by rotating the fullerene chains by $102^\circ$ with respect to their arrangement in KC$_{60}$.
As expected, at this stage, the diffraction profile was not well described during the Rietveld refinement ($R_\mathrm{wp}= 7.68\%$). However, a significant improvement was observed when the positions of the carbons C(15), which are involved in the \textit{single} C-C bonds, and their neighbours (C(5) and C(16)) were refined. Although the refinement process was always rather stable, due to the high number of observables available (1401), soft constraints were applied to carbon positions of the whole asymmetric unit and the convergence was reached by slowly decreasing the F-factor (penalty factor) at each cycle.

Further enhancement of the fit (to $R_\mathrm{wp}= 5\%$) was gained with the refinement of the carbon C(1) forming the four-membered carbon ring and its neighbours (C(2) and C(3)). Finally, the fractional coordinates of all the remaining ten C atoms of the asymmetric unit were refined stepwise. The best agreement between the observed and calculated data was obtained for an interfullerene C(15)-C(15) bond of 1.61(2) \AA. The latter distance is significantly smaller than that previously derived from synchrotron data (d$_{C(15)-C(15)}$ = 1.75(2)\AA \cite{margadonna04}), falling into a more reasonable range if one compares it with a typical C-C bond length. Moreover, no anomalous distances were detected in the fullerene molecule, the bond lengths being close to the standard values of 1.54 \AA\ and 1.42 \AA, for sp$^{3}$ and sp$^{2}$ C-C bonds respectively.

\begin{figure}
\includegraphics[width=0.9\textwidth]{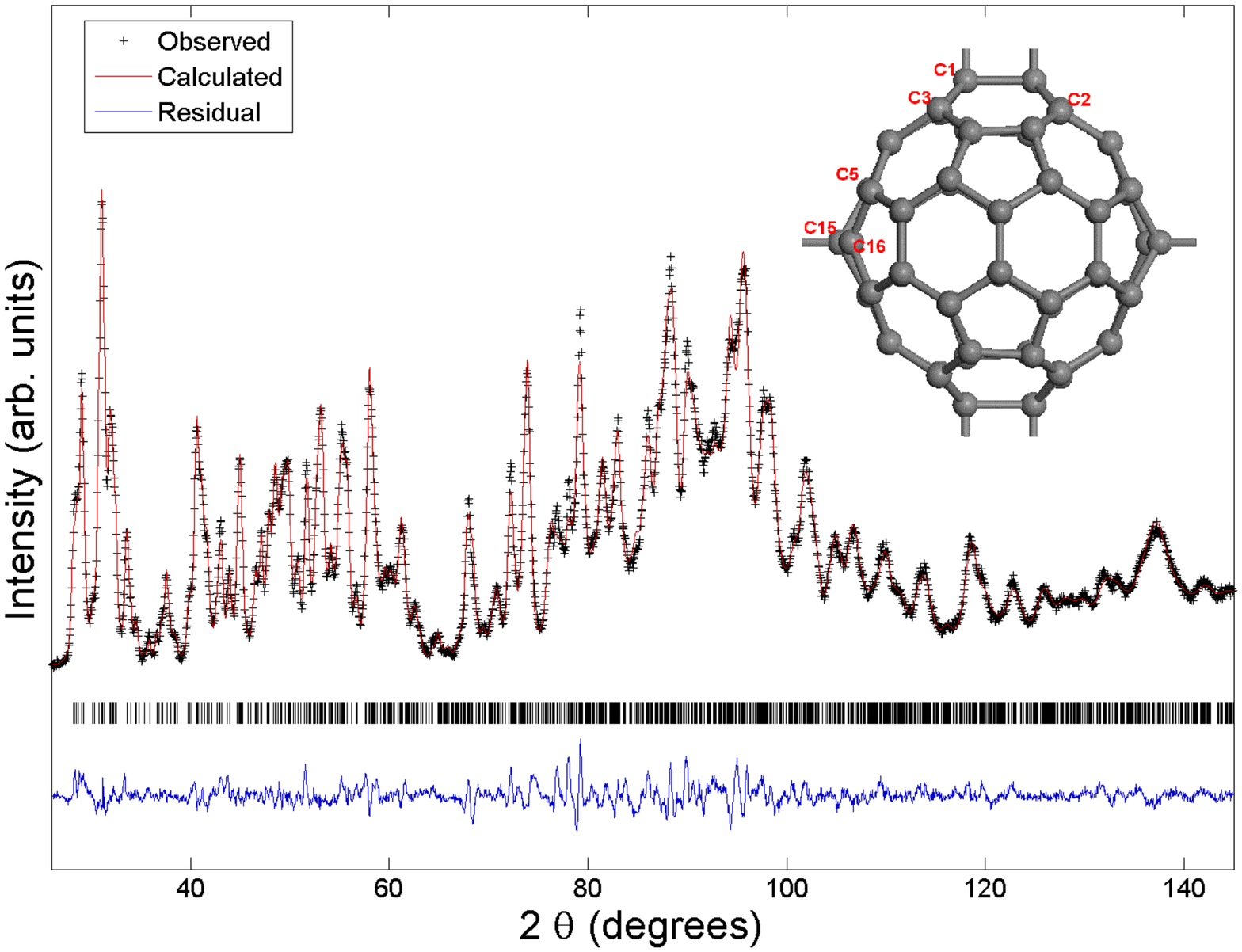}
\caption{Powder diffraction pattern of $^{7}$Li$_4$C$_{60}$ at 3.5 K. The data are indicated by black  crosses, the Rietveld fit by the solid red line and the difference by the solid blue line. Ticks marks indicate the reflection positions ($R_\mathrm{wp}$ = 3.83\%, $R_\mathrm{F2}$ = 1.16\%). The fulleride structure is displayed in the inset, with the carbon nomenclature.}
\label{fig:1-Rietveld}
\end{figure}

Solid State NMR and DC and AC conductivity \cite{arcon08} clearly indicate that the Li ions are pinned at special positions in the lattice at the temperature at which the diffraction data were collected. This means that no ionic diffusion is expected at 3.5 K. The structural analysis performed at room temperature using the synchrotron data distinguished two types of symmetry inequivalent Li ions, located respectively inside the pseudo-tetrahedral (Li$_T$) and pseudo-octahedral (Li$_O$) voids of the parent cubic lattice \cite{margadonna04}. In the present, low temperature neutron investigation, the study of the small alkali ion position was attempted using Fourier map analysis, \textit{i.e.} exploiting the large contrast arising from the negative neutron scattering length of $^{7}$Li. We found that a significant fraction of intercalated alkali ions occupy the position (0.44(1) 0 0.76(1)), which corresponds to the small pseudo-tetrahedral voids, in agreement with the previous analysis. The insertion of Li$_T$ atoms with full occupancy improved significantly the quality of the fit ($R_\mathrm{wp}$ = 3.99\%).\\
In contrast, the Fourier maps did not provide clear evidence of the occupation of the larger pseudo-octahedral sites, showing at these positions an unexpectedly poor contrast. The reason for this observation was first ascribed to the combined effect of the small absolute value of $^{7}$Li scattering length, as compared to C, in addition to the low amount of the alkali ions in the compound. In fact, if both effects are taken into account, the contribution of Li to the scattered intensity turns out to be two orders of magnitude smaller than that of carbon. Another hypothesis is that the Li$_O$ ions present a certain degree of disorder, \textit{e.g.} populating different octahedral sites in the lattice with similar energy and experiencing slightly different local potentials. This hypothesis is supported by DFT calculations \cite{ricco09}. The spread of Li$_O$ ions over different positions in the octahedral voids in the lattice is also expected to strongly reduce the contrast of the Fourier maps.\\
However, a careful analysis of the scattering reveals the existence of negative intensity close to the position (0.28, 0, 0.32), which is significantly shifted with respect to the center of the octahedral void. The insertion of the Li$_O$ alkali ion at this position with a full occupancy further improved the agreement factor ($R_\mathrm{wp}$= 3.83\%), suggesting that Li$_O$ ions tend to move away from the center of the octahedral sites with decreasing temperature. A similar behaviour was recently observed for highly-doped Li fullerene compounds, where Li ions were found to form clusters around the center of the tetrahedral positions of the parent cubic lattice \cite{giglio14} in the ground-state, leaving the octahedral sites unoccupied. \\
In the final configuration, the closest Li-Li distance found is of $2.9(1)$ \AA, similar to the atomic distance found in Li metal ($3.04$ \AA), while the Li$^+$-C$_{60}$ contacts reveal a short distance of the order of $\sim 2.4$ \AA, which is equal to the sum of the ionic radius of Li$^+$ ($0.7$ \AA) and the Van der Waals radius of C ($1.7$ \AA). The result of the Rietveld refinement performed on the neutron data is displayed in Fig.~\ref{fig:1-Rietveld}.

\subsection{Lattice Dynamics Calculations}
\subsubsection{Geometry optimization and electronic properties}

\begin{figure}[ht!]
     \begin{center}
        \subfigure[]{%
           \label{fig:Li4C60_I2m_4cells}
           \includegraphics[width=0.5\textwidth]{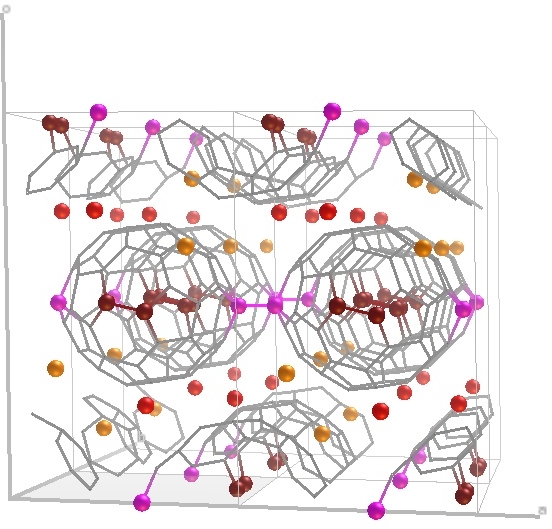}
        }
         \subfigure[]{%
            \label{fig:Li4C60_I2m_4cells_1plane}
            \includegraphics[width=0.5\textwidth]{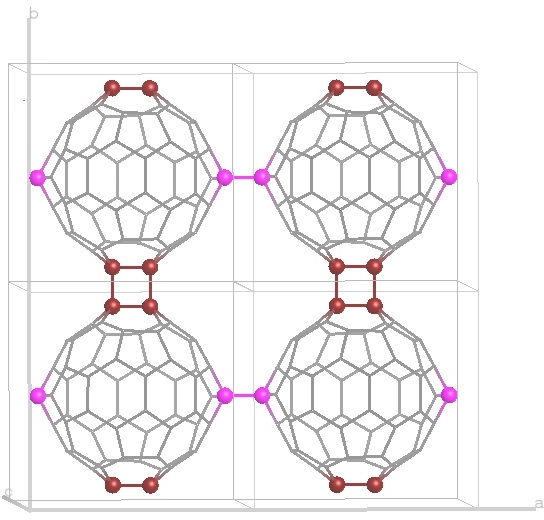}
        }%
    \end{center}
    \caption{%
        a) Structure of Li$_4$C$_{60}$ as obtained from geometry optimisation, in the space group I2/m. Grey sticks are carbon bonds, magenta and brown spheres are carbon atoms involved in the covalent intermolecular bonds: [2+2] bridges and single bonds respectively. Red and  yellow spheres represent Li$_T$ and Li$_O$ respectively. b) View along the $\vec c$ axis of one polymeric plane (the Li ions have been omitted for clarity).)
     }%
   \label{fig:Li4C60_I2mcell}
   
\end{figure}

The geometry relaxation was performed in two steps: the first one consisted of optimizing the atomic positions while keeping the lattice parameters at the values obtained by neutron diffraction. The experimental structure described above was used as the starting structure. The optimized geometry obtained this way was further relaxed by allowing lattice parameters to vary as well. The optimized structure was then considered for phonon calculations. We have checked that both relaxed geometries gave comparable results in terms of phonon dispersion curves and phonon density of states. The full relaxation however gave a better result in terms of there being no negative phonon frequencies in any part of the Brillouin zone and a slightly better agreement with the neutron derived phonon DOS.
The resulting structure has lattice parameters $a$ = 9.38 \AA, $b$ = 9.12 \AA, $c$ = 15.59 \AA, with angles $\alpha$ = $\gamma$ = 90$^\circ$ and monoclinic angle $\beta$ = 91.18$^\circ$. This gives a  $\Delta V / V$ = 5\% expansion of the lattice with regards to the experimental structure. Relaxed interatomic distances are, $a_{cc}(66)\sim$ 1.44 \AA\ (sp$^2$ bonds between two hexagons); $a_{cc}(65)\sim$ 1.42 \AA\ (sp$^2$ bonds between one hexagon and one pentagon); $a_{cc}(sb)\sim$ 1.56 \AA\ (sp$^3$ single bond between adjacent molecules), $a_{cc}([2+2])\sim$ 1.64 \AA\ (sp$^3$ bond in the [2+2] cycloaddition). In agreement with the diffraction investigations, the Li atoms are substantially displaced with regards to the centers of the parent tetrahedral and octahedral voids in cubic $C_{60}$. They form atomic planes of atoms in the space between two successive polymer sheets (see Fig.~\ref{fig:Li4C60_I2mcell}).

\subsubsection{Raman and IR modes}

\begin{figure}[ht!]
     \begin{center}
        \subfigure[B$_g$ Raman active mode at 102.5 meV (827 cm$^{-1}$)]{%
           \label{fig:R_102meV}
           \includegraphics[width=0.45\textwidth]{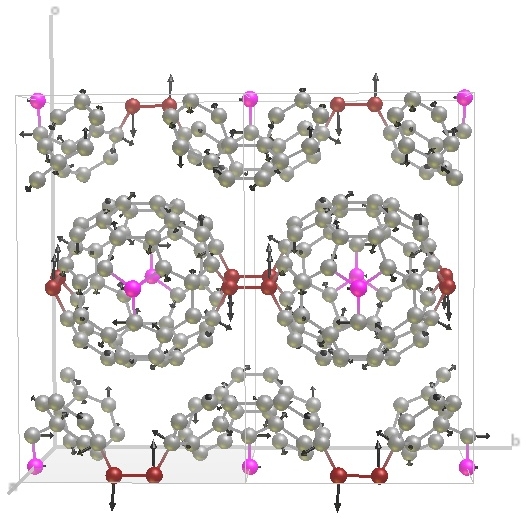}
        }
        \subfigure[A$_g$ Raman active mode at 104.7 meV (845 cm$^{-1}$)]{%
           \label{fig:R_105meV}
           \includegraphics[width=0.45\textwidth]{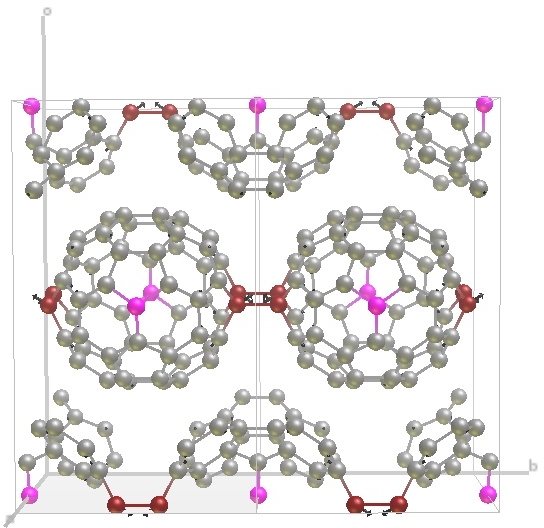}
        }\\
         \subfigure[A$_g$ Raman active mode at 123 meV (992 cm$^{-1}$)]{%
            \label{fig:R_123meV}
            \includegraphics[width=0.45\textwidth]{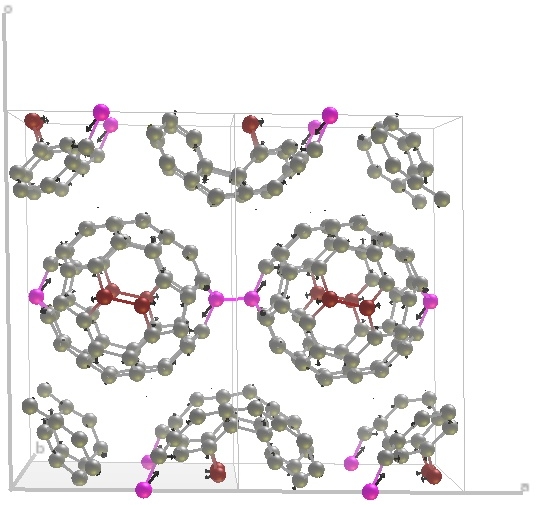}
        }
        \subfigure[A$_g$ Raman active mode at 178.6 meV (1441 cm$^{-1}$)]{%
            \label{fig:pinch_mode}
            \includegraphics[width=0.25\textwidth]{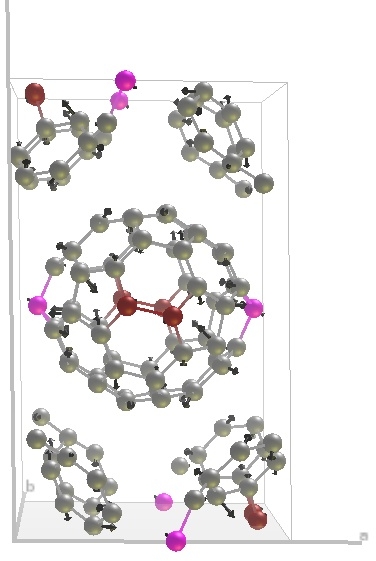}
        }%
    \end{center}
    \caption{%
        Selected Raman Active modes discussed in the text. The Li atoms are omitted for clarity. 
     }%
   \label{fig:RamanModes}
   
\end{figure}

The crystal symmetry point group of Li$_4$C$_{60}$ is $C_{2h}$. The vibrational representation of the system is composed of the non-degenerate A$_g$(R), B$_g$(R), A$_u$(IR) and B$_u$(IR) irreducible representations. The phonon modes at point $\Gamma$ are either Raman (R) or Infrared (IR) active. The complete list of the modes, classified according to their symmetry and frequency, are presented in Tables ~\ref{tab:tab_ram_cdz} and ~\ref{tab:tab_ir_cdz}. The calculated frequencies are compared to those of the principal features observed in the Raman, IR and High Resolution Electron Energy Loss Spectroscopy (HREELS) spectra reported in the literature \cite{ricco05,wagberg04,yasukawa01,macovez2008}. Note that HREELS is subject to the same selection rules as IR. The association between the calculated and the observed modes is based on the best match between their frequency and other arguments like, for example, their relative position with regards to common spectral gaps.
Compared to bulk C$_{60}$, for which only 13 modes are Raman active and 6 are IR active \cite{dresselhaus} (including lattice modes), resulting in 19 observable modes, the reduced symmetry of the Li$_{4}$C$_{60}$ lattice, added to the deformation of the fullerene cages due to polymerization, increases the number of optically active modes to 95 Raman and 94 IR active modes. This renders the interpretation of the Raman and IR spectra based only on the  frequency match between the experimental and calculated features very difficult. A definitive attribution of the observed features would require the calculation of the intensity of the Raman/IR active modes which is beyond the scope of this paper.\\
In this section we restrict the discussion to two types of Raman active modes which are of particular interest and which are extensively discussed in the literature. The first one concerns the double peak feature located around 120 meV (119 and 121.5 meV in Li$_4$C$_{60}$) which is often associated to the stretching of the [2+2] bond in polymeric C$_{60}$. However its observation in the Raman spectrum of the single bonded polymer Na$_4$C$_{60}$ has raised a question about this attribution \cite{ricco05,wagberg04}. 
In the experimental Raman spectra, this feature appears well separated from the preceding Raman peak at lower frequency by a $\sim$15 meV gap. In Tab.~\ref{tab:tab_ram_cdz}, these two extreme features are observed at 119 and 102.7 meV respectively. Our calculations also predict a 15 meV frequency range populated essentially by IR active modes. They create a gap in the Raman spectra between 108 meV and 123 meV. Therefore, we attribute the observed 119 meV features to a A$_g$ active mode which is represented in Fig.~\ref{fig:R_123meV}. This mode shows a complex displacement pattern, composed mostly of single-bond deformations, but also of [2+2] contributions. In contrast, the modes located at lower frequencies imply essentially movements of the atoms forming the [2+2] bridge. In Fig.~\ref{fig:R_102meV} we show the B$_g$ mode calculated at 102.5 meV. Its displacement pattern represents the buckling of the [2+2] square-like bridge out of the polymer plane, \textit{i.e} along the $\vec c$ direction. We attribute to this mode the peak observed at 102.7 meV in the Raman spectra. Other Raman active modes, implying the deformation of this intermolecular bridge, are calculated in this frequency range: a B$_g$ mode at 103.8 meV and a A$_g$ mode at 104.7 meV, the latter mode being associated with the stretching of the bridge along the $\vec b$ direction (see Fig.~\ref{fig:R_105meV}). They could also be attributed to the feature observed in the Raman spectra at 102.7 meV. 

In conclusion, our calculations propose that the features in the Raman spectra observed in the range around 105 meV are associated to the [2+2] bond deformations, \textit{e.g.} at frequencies significantly smaller than those of polymeric forms of C$_{60}$ obtained at high pressure and high temperature.
Their contribution to the phonon density of states (DOS) is sharp and observed at $\sim$103-105 meV in the total DOS spectrum shown in Fig.~\ref{fig:pdos}. The deformation of the [2+2] cyclic bond, therefore appears as a highly localised mode featuring a sharp peak in the pDOS of the C atoms involved in this bond (see brown line in the top part of Fig.~\ref{fig:pdos}). 

In contrast, the phonon modes involving single bond contributions are calculated to be hybridized with other molecule deformations to a large extent. They give a significant contribution in the 120-150 meV range. Their contribution to the DOS is relatively smooth (see magenta line in the top part of Fig.~\ref{fig:pdos} top).

The second Raman feature that we discuss in this section concerns the so--called "pentagonal pinch mode". It refers to a C$_{60}$ mode, of A$_g$ symmetry in bulk C$_{60}$, located at 182 meV (1469 cm$^{-1}$) involving tangential displacements of the atoms and in particular, contraction of the pentagonal rings of the fullerene. Its frequency is known to red-shift depending on the number of polymeric bonds per molecule \cite{wagberg99,davydov00} and/or the charge carried by the fullerene \cite{duclos1991} according to the empirical rule: -0.8 meV per excess electron, -0.3 meV per single covalent bond and -0.7 meV per [2+2] covalent bond.
A mode in the Raman spectra of Li$_4$C$_{60}$ is observed in this frequency range (see Tab.~\ref{tab:tab_ram_cdz}) at a frequency of 178.8 meV, \textit{i.e.} red-shifted by $\sim$3 meV with regards to the bulk value. Our calculation predicts a A$_g$ Raman active mode at a frequency of 178.6 meV. This pure cage deformation mode is represented in Fig.~\ref{fig:pinch_mode} and presents a pattern close to the one expected from the A$_g$(2) "pentagonal pinch mode" of bulk C$_{60}$. Following the empirical rule described above, we anticipate a charge transfer of about 2e per cage \cite{ricco05,pontiroli06}, a value substantially lower than that proposed by Macovez \textit{et al.} who  derived a value closer to 3e (2.7e) from the analysis of the photoemission spectrum of a Li$_{4}$C$_{60}$ film \cite{macovez2008}. Both predictions however agree on the partial nature of the charge transferred from the lithium to the fullerene cage, a tendency already known in intercalated fullerene systems \cite{Tomaselli}.
The analysis of our simulations, using the Bader \cite{bader,tang09} approach, indicates a total charge of 726.1 electrons per C$_{60}$ which means a charge of $\sim$-3e per cage and a partial charge transfer from the Li atoms to the C$_{60}$ cage which amounts to $\sim$0.75e per alkali ion. This value is in very good agreement with the value proposed by Macovez \textit{et al.}. In addition, the calculations reproduce well the experimental frequency of the pentagonal pinch mode Raman feature. This observation reinforces the difficulty of deriving the excess of charge transferred to the fullerene in polymeric form based on the frequency of the Raman feature alone. Any use of the above described rule will lead to an underestimate of the on--ball charge transfer. 

\begin{table}[h!]
\begin{tabular}{*{2}{p{1cm}}|*{2}{p{1cm}}|*{2}{p{1cm}}||*{2}{p{1cm}}|*{2}{p{1cm}}|*{2}{p{1cm}}*{2}{p{1cm}}} 
%
\multicolumn{6}{c||}{A$_g$} & \multicolumn{6}{c}{B$_g$} \\
Calc. & \textit{Exp.} & Calc. & \textit{Exp.} & Calc. & \textit{Exp.} & Calc. & \textit{Exp.} & Calc. & \textit{Exp.} & Calc. & \textit{Exp.} \\

\specialrule{.2em}{.1em}{.1em}


11.0 &   & 14.9 &   & 16.9 & \textit{16.5} & 18.0 &   & 19.1 & \textit{19.6} & 24.3 & \textit{24.2} \\
30.2 &   & 32.6 &   & 36.5 &   & 25.7 &   & 32.3 &   & 35.0 & \textit{34.2} \\
39.9 &   & 44.4 & \textit{44.4} & 51.1 & \textit{52.1} & 46.8 &   & 50.2 & \textit{50.7} & 54.9 &   \\
53.9 & \textit{52.9} & 56.6 & \textit{56.5} & 59.6 & \textit{60.1} & 58.6 &   & 62.2 &   & 67.6 &   \\
62.0 & \textit{62} & 65.5 &   & 66.4 & \textit{67.1} & 68.7 &   & 71.1 &   & 74.3 & \textit{74.4} \\
70.1 &   & 71.6 &   & 78.1 &   & 75.5 & \textit{75.5} & 79.9 &   & 82.3 & \textit{81.5} \\
81.9 & \textit{80.5} & 84.2 & \textit{83.4} & 84.5 & \textit{85.3} & 86.8 &   & 89.3 &   & 91.8 & \textit{93} \\
86.4 & \textit{86.2} & 90.3 & \textit{90.3} & 91.7 &   & 95.3 & \textit{94.3} & 97.3 & \textit{97} & 97.7 &   \\
95.6 &   & 98.9 &   & 101.5 & \textit{100.3} & 98.8 &   & 102.5 & \textit{102.7} & 103.8 &   \\
104.7 &   & 123.0 & \textit{119} & 128.2 & \textit{121.5} & 108.0 &   & 128.1 & \textit{121.5} & 132.3 &  \textit{130.4} \\
131.0 & \textit{129.3} & 134.4 & \textit{134.1} & 136.7 & \textit{137} & 134.9 & \textit{135.5} & 138.8 &   & 144.6 &   \\
138.5 &   & 141.9 &   & 144.6 & \textit{145.7} & 146.8 &   & 152.3 & \textit{151.6} & 157.3 & \textit{157.7} \\
147.6 & \textit{148.5} & 155.6 &   & 159.9 &   & 160.2 &   & 162.7 & \textit{163.5} & 165.4 &   \\
161.1 &   & 162.0 & \textit{161.8} & 165.9 &   & 167.6 & \textit{167.4} & 169.2 &   & 173.0 & \textit{172.1} \\
167.6 &   & 170.1 & \textit{169.8} & 173.7 &   & 177.9 &   & 184.6 & \textit{184.2} & 185.7 &   \\
175.8 &   & 178.6 & \textit{178.8} & 184.7 &   & 189.0 & \textit{195.6} &  &  &  &  \\
184.7 &   & 186.5 & \textit{187.2} &  &  &  &  &  &  &  &  \\

\end{tabular} 
\caption{ \label{tab:tab_ram_cdz} Raman active modes frequencies (in meV) calculated for A$_g$ (\textit{Left}) and B$_g$ (\textit{Right}) symmetry. The frequencies of the 45 discernible features observed in the experimental Raman spectra obtained at room temperature using a red (1.96 eV) laser excitation on a Li$_4$C$_{60}$ polymer sample \cite{ricco05} (see also Ref.~\onlinecite{wagberg04}) are also reported. The experimental frequencies (\textit{in italic}) are placed inside the \textit{Exp.} columns at positions in the table the closest to those of modes predicted by the simulations.}
\end{table}

\begin{table}[h!]
\begin{tabular}{*{2}{p{1cm}}|*{2}{p{1cm}}|*{2}{p{1cm}}||*{2}{p{1cm}}|*{2}{p{1cm}}|*{2}{p{1cm}}*{2}{p{1cm}}} 
\multicolumn{6}{c||}{A$_u$} & \multicolumn{6}{c}{B$_u$} \\
Calc. & \textit{Exp.} & Calc. & \textit{Exp.} & Calc. & \textit{Exp.} & Calc. & \textit{Exp.} & Calc. & \textit{Exp.} & Calc. & \textit{Exp.} \\

\specialrule{.2em}{.1em}{.1em}


18.9 &   & 25.8 & \textit{26$^{\dag}$} & 38.0 & \textit{38$^{\dag}$} & 11.5 &   & 15.0 & \textit{14$^{\dag}$} & 35.1 &   \\
39.7 &   & 44.9 &   & 46.2 &   & 38.5 &   & 41.9 &   & 45.2 &   \\
47.2 &   & 48.2 & \textit{48$^{\dag}$} & 57.3 & \textit{56$^{\ast}$} & 46.3 &   & 47.5 &   & 48.6 &   \\
61.0 &   & 63.7 &   & 65.6 & \textit{65$^{\ast\dag}$} & 49.0 &   & 59.4 &   & 59.6 &   \\
68.1 &   & 78.5 &   & 79.8 &   & 66.1 & & 66.8 &   & 69.6 &   \\
80.7 & \textit{81.5$^{\ast}$} & 82.6 &   & 83.3 &   & 72.0 & \textit{71$^{\ast\dag}$} & 78.4 & \textit{76$^{\ast}$} & 79.1 &   \\
86.1 &   & 88.4 &   & 89.6 &   & 80.6 &   & 84.1 &   & 86.1 &   \\
92.0 &   & 93.4 & \textit{94$^{\ast}$} & 94.6 & \textit{96$^{\ast}$} & 88.4 &   & 89.0 &   & 90.8 & \textit{91$^{\ast}$} \\
108.7 &   & 115.8 &   & 118.3 &   & 98.2 & \textit{99$^{\ast\dag}$} & 104.2 &   & 111.0 &   \\
120.2 &   & 130.1 &   & 133.2 &   & 115.7 &   & 116.2 &   & 119.3 &   \\
143.2 &   & 147.1 & \textit{147$^{\ast\dag}$} & 151.1 & \textit{151$^{\ast}$} & 135.1 &   & 137.9 &   & 140.9 &   \\
154.2 &   & 157.2 &   & 159.7 &   & 143.2 &   & 148.8 &   & 152.7 &   \\
165.2 &   & 166.4 &   & 168.2 &   & 155.3 & \textit{155.5$^{\ast}$} & 162.0 &   & 164.2 &   \\
171.6 &   & 174.8 &   & 177.2 & \textit{177$^{\dag}$} & 164.9 &   & 166.8 &   & 170.1 &   \\
185.8 &   & 186.6 &   & 188.9 &   & 174.0 &   & 175.6 &   & 179.1 &   \\
 &  &  &  &  &  & 181.9 &   & 185.3 &   & 186.6 &   \\

\end{tabular} 
\caption{ \label{tab:tab_ir_cdz} \textit{Left}: 
IR and HREELS active modes frequencies (in meV) calculated for A$_u$ (\textit{Left}) and B$_u$ (\textit{Right}) symmetry. Frequencies of the principal features observed in the experimental IR (from Ref.~\onlinecite{yasukawa01}, symbol $^{\ast}$) and from HREELS spectra (from Ref.~\onlinecite{macovez2008}, symbol $^{\dag}$) are also reported. The experimental frequencies (\textit{in italic}) are placed inside the \textit{Exp.} columns at position in the table the closest to those of modes predicted by the simulations.}
\end{table}

\subsubsection{Phonon dispersion curves and Phonon DOS}

\begin{figure}
\includegraphics[width=1.\textwidth]{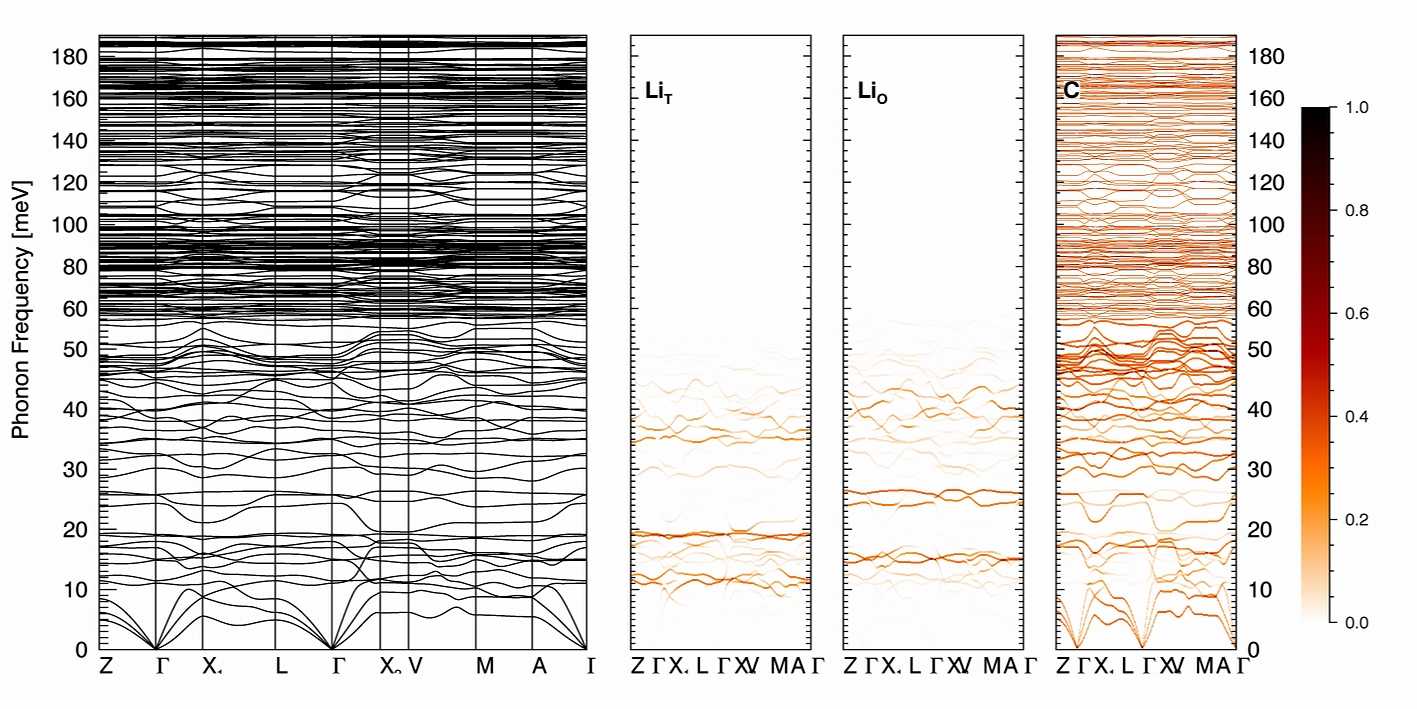}
\caption{Left: Dispersion curves along high symmetry directions. Right: Partial dispersion curves along high symmetry directions. From left to right, contributions from Li$_T$, Li$_O$ and carbon atoms. The color scheme represents the value of the participation factor of the atom in the modes: the darker the color, the higher the participation (normalized to unity, \textit{i.e.} a value of one represents maximum participation factor).}
\label{fig:disp_curves_all}
\end{figure}

Figure \ref{fig:disp_curves_all} shows the dispersion curves of the phonon branches calculated along several high symmetry directions in the Brillouin zone. 
The high symmetry point coordinates are given in table \ref{tab:high_sym_pts_coord} in the system of reciprocal vectors ($\vec{A},\vec{B}, \vec{C}$) of the crystallographic unit cell (I2/m). Due to the weak monoclinic character of the unit cell, propagation vectors of the form [0, 0, $\xi$] will have major out--of--plane components and refer to phonon modes propagating out of the polymer plane. In contrast, propagation vectors of the form [$\xi$, $\eta$, 0] refer to phonon modes propagating in the polymer plane.
\begin{table}[h!]
\begin{tabular}{c|c|c|c|c|c|c|c} 
Label & [Q$_1$,Q$_2$,Q$_3$] & Label & [Q$_1$,Q$_2$,Q$_3$] & Label & [Q$_1$,Q$_2$,Q$_3$] & Label & [Q$_1$,Q$_2$,Q$_3$] \\
\hline
$\Gamma$ & [0,0,0] & X$_1$ & [$\frac{1}{2}$,0,0] & X$_2$ & [0,$\frac{1}{2}$,0] & L & [$\frac{1}{2}$,$\frac{1}{2}$,0] \\
Z & [0,0,1] & V & [0,$\frac{1}{2}$,$\frac{1}{2}$] & M & [$\frac{1}{2}$,0,$\frac{1}{2}$] & A & [$\frac{1}{2}$,0,-$\frac{1}{2}$]
\end{tabular}
\caption{ \label{tab:high_sym_pts_coord} High symmetry point coordinates expressed in the system of reciprocal vectors ($\vec{A}$, $\vec{B}$, $\vec{C}$) of the crystallographic unit cell (I2/m). Note that due to the weak monoclinic ($\beta$ = 91.18$^\circ$) character of the unit cell, the vectors $\vec{A}$, $\vec{B}$ and $\vec{C} $ are close to being collinear to the single-bonds, [2+2] bonds and normal to the polymer plane directions respectively.}
\end{table}

\begin{figure}
\includegraphics[width=1.\textwidth]{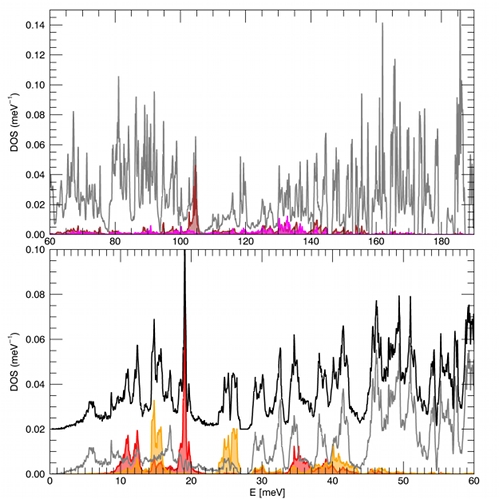}
\caption{Total (black line) and partial density of state of the different atomic species: all C (grey line), Li$_T$ (red line and area), Li$_O$ ( yellow line and area), C(1) --involved in the [2+2] bonds -- (brown line and area), C(15) --involved in the single intermolecular bonds-- (magenta line and area). The total density of state was shifted by 0.02 meV$^{-1}$ for clarity in the low frequency part (bottom curves) of the spectrum. The total DOS is not represented in the high frequency range (top curves), as it is totally dominated by the C partial DOS in this range. The C(1) and C(15) partial DOS are only shown in the high frequency range.}
\label{fig:pdos}
\end{figure}

The dispersion curve pattern is typical of an anisotropic molecular crystal, its molecular nature being revealed by flat "molecular" modes in the high frequency range (E $\ge \sim$ 30 meV) and the presence of a small 2 meV gap between 26.5 and 28.5 meV. The strong anisotropy between the in--plane covalent bonds and the out-of-plane van der Waals interactions is observed by the rather flat dependence of the molecular modes along the \textit{e.g.} $\Gamma - Z$ line, while a substantially larger dispersion is calculated for \textit{e.g.} the modes in the [110, 120 meV] range along the $\Gamma - X_1$ direction. These high frequency molecular modes essentially involve carbon displacements, Li ions being largely immobile, as seen from the weak Li participation factors in this energy range (see Fig.\ref{fig:disp_curves_all} Li$_O$ and Li$_T$) and the Li partial density of states (pDOS) (see Fig.\ref{fig:pdos}). As already discussed for other polymeric forms of C$_{60}$, these phonon modes share strong similarities with those of bulk C$_{60}$, but the DOS spectrum also differs significantly in some places, which can be ascribed to the reduced symmetry of the polymeric cage and to the different nature of the hybridization of some C-C bonds at the cage surface \cite{pintchovius96, kolesnikov96, schober97}. 

The 2D character of the polymeric sheet is reflected into the acoustic region of the dispersion curves. In particular, two of the three acoustic branches have a steep dispersion in the $\Gamma - X_2$ direction, corresponding to displacements of the cages in the polymer plane. The third acoustic branch, on the other hand, shows a much flatter dependence with $q$ close to $\Gamma$ and presents a significant up-turn for the $\Gamma - X_1$ and $\Gamma - X_2$ direction with increasing $q$. This general feature, typical of a 2D system, is well known from graphite \cite{nicklow72} and has also been calculated for polymeric Na$_4$C$_{60}$ \cite{schober99}.

\begin{figure}[ht!]
     \begin{center}
     
        \subfigure[ Libration around the $\vec c$ axis at 25.7 meV (B$_g$(4))]{%
            \label{fig:model1}
            \includegraphics[scale=0.37]{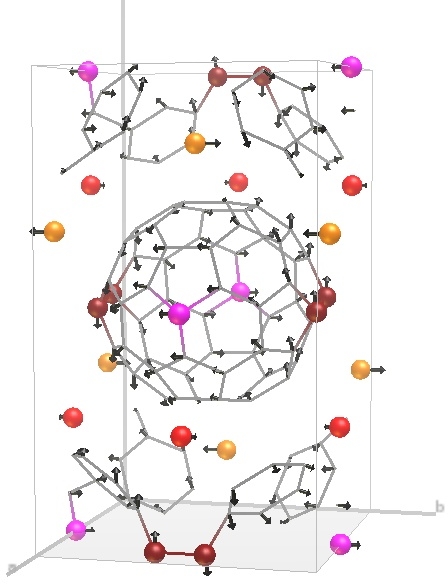}
        }%
        \subfigure[ Libration around the $\vec a$ axis at 32.3 meV (B$_g$(5))]{%
            \label{fig:model2}
            \includegraphics[scale=0.37]{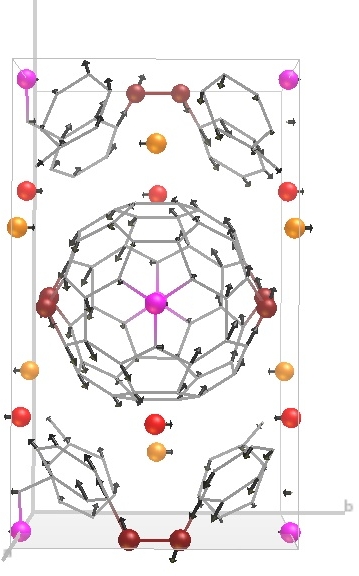}
        }%
        \subfigure[ Libration around the $\vec b$ axis at 32.6 meV (A$_g$(5))]{%
            \label{fig:model3}
            \includegraphics[scale=0.37]{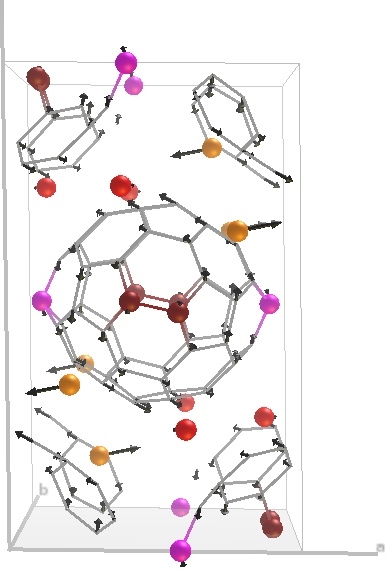}
        }%
    \end{center}
    \caption{%
        The three rotation--like modes and their symmetry.
     }%
   \label{fig:ModesGammaLib}
\end{figure}

\begin{figure}[ht!]
     \begin{center}
            \subfigure[ "$H_g$ ghost" mode at 16.9 meV (A$_g$(3))]{%
            \label{fig:modeH1}
            \includegraphics[scale=0.37]{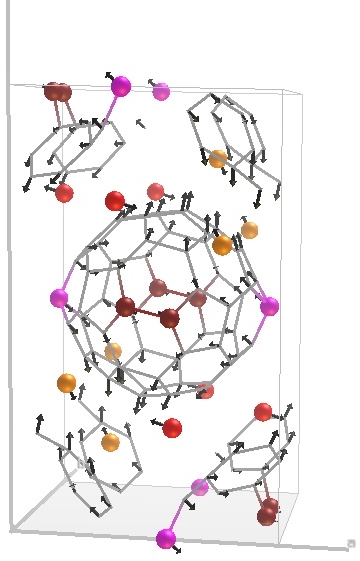}
        }%
        \subfigure[ "Hybrid mode" at 18 meV (B$_g$(1))]{%
            \label{fig:modeH2}
            \includegraphics[scale=0.37]{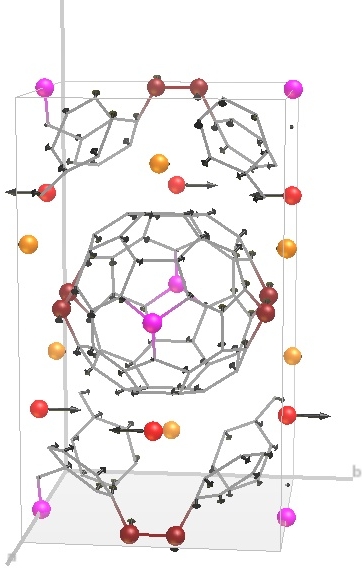}
        }%
        \subfigure[ "Hybrid mode" at 30.2 meV (A$_g$(4))]{%
            \label{fig:modeH3}
            \includegraphics[scale=0.37]{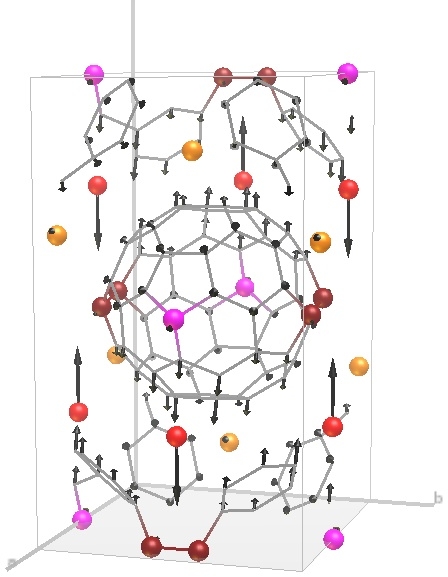}
        }%
    \end{center}
    \caption{%
        Selected "H$_g$ Ghost" and "Hybrid" modes and their symmetry.
     }%
   \label{fig:ModesGammaVib1}
\end{figure}

The anisotropy and reduced symmetry of the system is also clearly seen in the lattice mode region (E $\lesssim$ 15 meV) of the DOS represented in the bottom part of Fig.\ref{fig:pdos}. The flat contribution of the out--of--plane transverse acoustic mode along the ($X_2VMA$) directions gives the first van Hove singularity at 5.5 meV. The anisotropy of the polymeric system is further reflected by the pseudo gap (4 meV) separating this contribution from the second van Hove singularity in the DOS observed at $\sim$ 9 meV, which results from the flat dependence of the in-plane acoustic modes in the same region of the Brillouin zone.\\
At higher frequencies, a series of flat optical branches is observed in the $\sim$ 10--20 meV range, followed by a pseudo gap in the range $\sim$ 22 meV and an intense flat band at $\sim$ 25 meV. Then follows the "real" gap of 2 meV between 26.5 meV and 28.5 meV featuring no intensity in the DOS spectrum. It is well known that the polymeric forms of C$_{60}$ have characteristic modes that appear in the "Gap" region of bulk C$_{60}$ ([8-33 meV], see \textit{e.g.} \cite{kolesnikov96,schober97,rols03} and references therein). In particular, cage libration modes, with the C$_{60}$ cages rotating around their principal axis of symmetry, \textit{i.e.} implying inter-molecular bonds to be slightly deformed, are usually found in this range. These modes, located at 3.5 meV in bulk monomeric C$_{60}$, are shifted to the 10--30 meV range for polymeric forms of C$_{60}$, depending on the precise bonding scheme. These librations are calculated at 25.7 meV (B$_g$(4)), 32.3 meV (B$_g$(5)) and 32.6 meV (A$_g$(5)) for Li$_4$C$_{60}$. Their rotation axes main lie along the $\vec{c}$, $\vec{a}$ and $\vec{b}$ unit cell vectors respectively, as represented in Fig.~\ref{fig:ModesGammaLib}. In monomeric $C_{60}$, the first intramolecular vibration is a Raman active five fold degenerate H$_g$ mode at 33 meV. Polymerisation lifts its degeneracy and we calculate "ghosts" of this mode distributed over a relatively large number of modes in the 10--30 meV frequency range, some involving a significant contribution of Li displacements, \textit{i.e.} hybridized with Li modes. An example of such a mode is shown in Fig.\ref{fig:modeH1} which represents a Raman active mode at 16.9 meV (A$_g$(3)). Tab.~\ref{tab:tab_ram_cdz} indicates that a feature in the Raman spectrum of the polymer is observed at 16.5 meV. All the optical branches mentioned above have flat dispersion, and they give sharp contributions to the DOS. In particular, the flat dispersion of the cage deformation mode in the 17 meV region (see Fig.\ref{fig:disp_curves_all}, right panel) gives a rather intense and sharp feature in the total DOS and C pDOS spectra.
%
%
%

\begin{figure}[ht!]
     \begin{center}    
        \subfigure[ Pure Li$_T$ at 11 meV (A$_g$(1))]{%
            \label{fig:modeLi1}
            \includegraphics[scale=0.37]{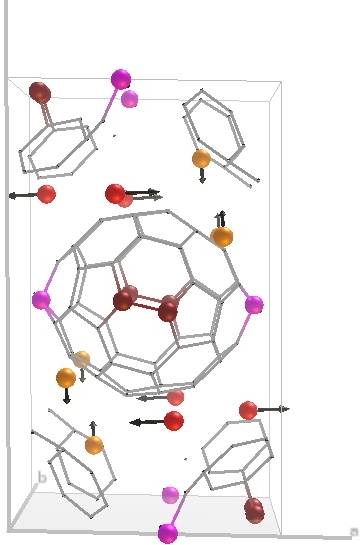}
        }%
        \subfigure[ Pure ferro Li$_T$ mode (translations along $\vec b$) at 18.9 meV (A$_u$(1))]{%
            \label{fig:modeLi2}
            \includegraphics[scale=0.37]{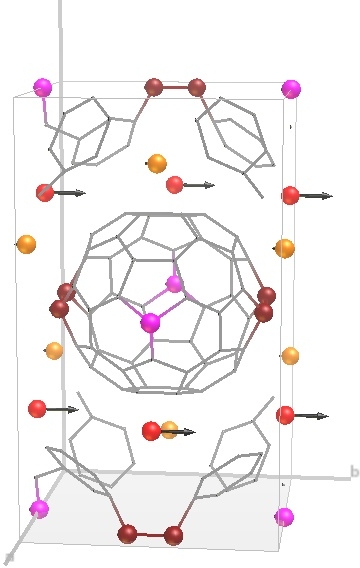}
        }%
        \subfigure[ Pure anti-ferro Li$_O$ mode (translations along $\vec b$) at 24.3 meV (B$_g$(3))]{%
            \label{fig:modeLi3}
            \includegraphics[scale=0.37]{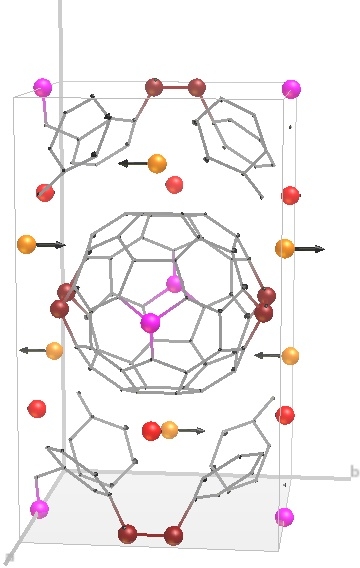}
        }%
    \end{center}
    \caption{%
        Selected pure Li modes and their symmetry.
     }%
   \label{fig:ModesGammaVib2}
\end{figure}

The Li phonon modes extend up to a maximum energy of $\sim$ 45 meV. Fig.\ref{fig:disp_curves_all} and Fig.\ref{fig:pdos} show that they are grouped into three energy bands for each Li ion type: centred around 12, 19 and 36 meV for Li$_T$ ions, and around 15, 25 and 40 meV for Li$_O$ ions. These groups imply either ferro (in--phase displacement of the two symmetry related ions) or anti-ferro translation modes (out--of--phase displacement of the two symmetry related ions) along the $\vec a$, $\vec b$ and $\vec c$ axis respectively (see Fig.~\ref{fig:ModesGammaVib2}). In the lowest frequency group, the modes concern both Li$_T$ and Li$_O$ ions, with concomitant orthogonal displacements along the $\vec a$ and $\vec c$ axis, as shown on Fig.~\ref{fig:modeLi1}.\\
The sharp contributions of these modes to the DOS, and the fact that the Li ions do not have a significant participation factor in the acoustic bands suggest a quite weak interaction between the Li and C atoms, \textit{i.e.} in good agreement with the partial charge transfer between the alkali atoms and the fullerene cages. The large mass difference between the Li and the fullerene explains the very weak coupling of the Li ion displacements to the polymer vibrations in the acoustic range, with a vanishing participation factor of the alkali atoms for non--zero propagation vectors (\textit{i.e.} except at the $\Gamma$ point). A careful analysis of the dispersion curves and polarization vectors reveals a certain degree of hybridisation between the fullerene modes and the Li$_O$ and Li$_T$ modes. In particular, it is found that the Li ions follow adiabatically the fullerene movements for the libration modes (see Fig.~\ref{fig:ModesGammaLib}). A coupling between the H$_g$ derived modes and the Li translations, either in--plane or out--of--plane is also observed (see \textit{e.g.} the modes shown in Fig.~\ref{fig:ModesGammaVib1}). These hybrid modes imply deformations of the fullerene cage and are of particular interest as they displace the Li ions far away from their equilibrium positions, invariably contributing to the large ionic diffusivity of this system.\\
A very sharp feature is observed in the Li$_T$ pDOS at a frequency of 19.2 meV, which originates from the dispersionless bands (see Fig.~\ref{fig:disp_curves_all}). No fullerene modes are calculated in this region over the entire Brillouin zone. In contrast the Li$_O$ related modes in the 25 meV range show a significant dispersion and a broader feature in the corresponding pDOS, which originates from the presence of fullerene modes in this region, \textit{e.g.} the libration around the $\vec a$ axis.

\subsubsection{Thermodynamic functions}
\label{sssec:thermo}

\begin{figure}
\includegraphics[width=0.9\textwidth]{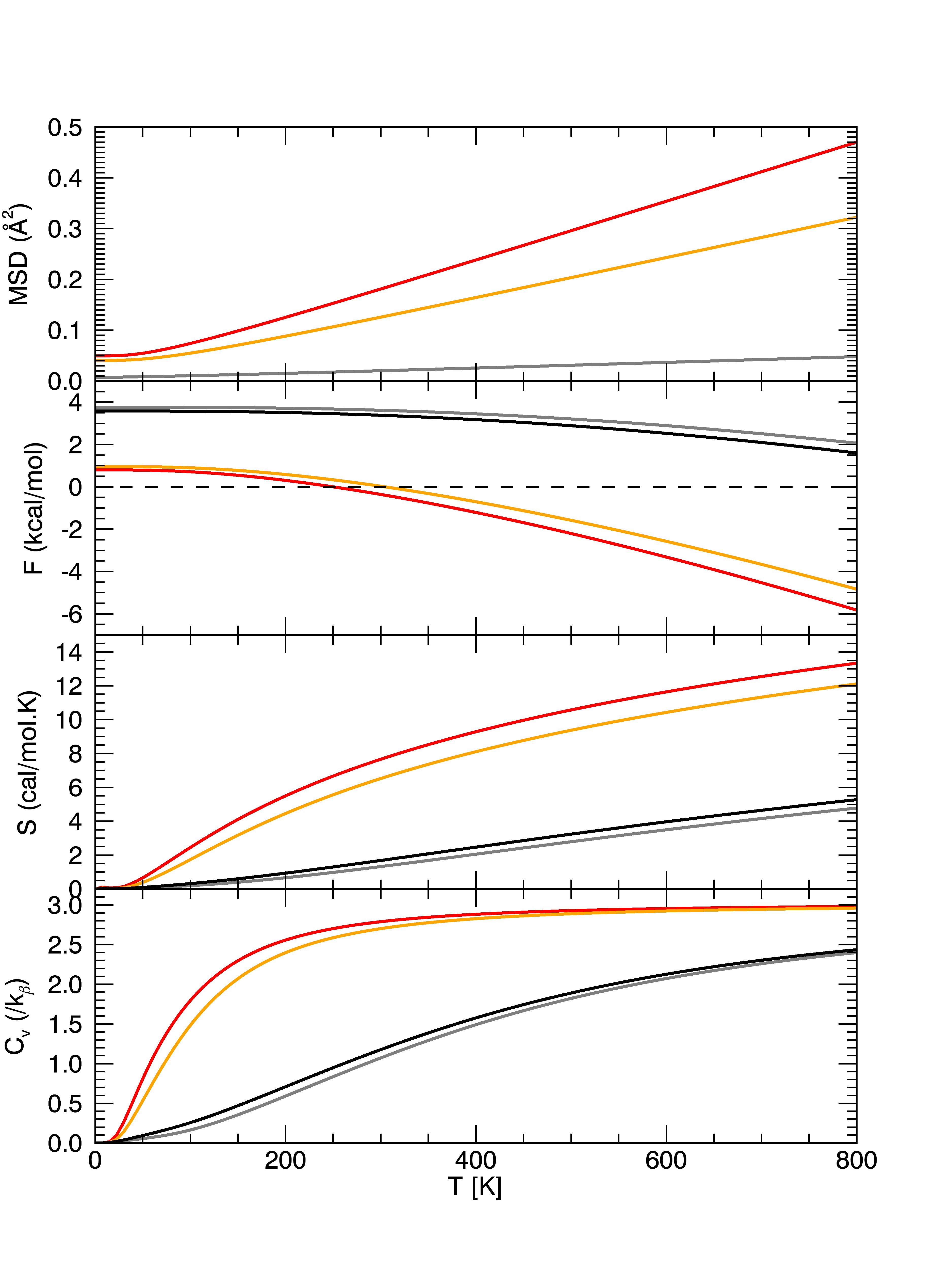}
\caption{From top to bottom: atomic mean squared displacement (MSD), molar free energy (F), molar vibrational entropy (S) and atomic heat capacity (C$_v$), for all atomic species (black) and for C (grey) Li$_T$ (red) and Li$_O$ ( yellow) atoms. $C_v$ is given per atom and in unit of $k_\beta$, the Boltzmann constant.}
\label{fig:thermod}
\end{figure}

As seen in Fig.\ref{fig:thermod}, the additive combination of low mass and frequency for Li vibrations is responsible for their very large mean square displacement (MSD) $\langle u^2 \rangle$, even at relatively low temperature. In particular, it is calculated to be much larger than the carbon atom equivalent in the temperature range from 0--800 K: $\langle u^2_{Li_T} \rangle$ (0 K) = $\langle u^2_{C} \rangle$ (800 K). The alkali sub--lattices appear therefore as soft lattices with large amplitude harmonic vibrations, embedded inside a stable carbon matrix, which shows a small degree of flexibility due to the soft cage deformation modes. The difference in stiffness and stability of the carbon and lithium sublattices is further reflected in the molar heat capacity and entropy represented in Fig.\ref{fig:thermod}: although the fraction of Li atoms in the system is weak, the specific heat at $\sim$ 100 K originates to a significant extent (close to a third) from Li vibrations, while the classical atomic limit of 3$k_\beta$ is almost reached at 300 K. By contrast, it is still half $k_\beta$ for carbon at 400 K. The large amplitude vibrations of Li also imply a large value of their corresponding vibrational entropy at relatively low temperatures, ($\sim$ 6 cal/mol.K at $\sim$ 200 K, \textit{i.e.} six times larger than that of carbon). The most interesting consequence of this is the Li derived free energy which becomes negative at a temperature of $\sim$ 250 K, suggesting that the Li sub-lattices become unstable above these temperatures, even in the harmonic approximation, while the fullerene framework retains its stability. If one includes anharmonic terms, the Li atom framework is likely to have melted at such temperatures. As no important change is expected for the global structure, the transition from an ordered Li sub-lattice to a liquid-like structure is certainly of second order character.

\subsection{Inelastic neutron scattering}
\label{ssec:INS}

\subsubsection{S(Q,$\omega$) maps of the polymer phase}
\begin{figure}
\includegraphics[width=1.\textwidth]{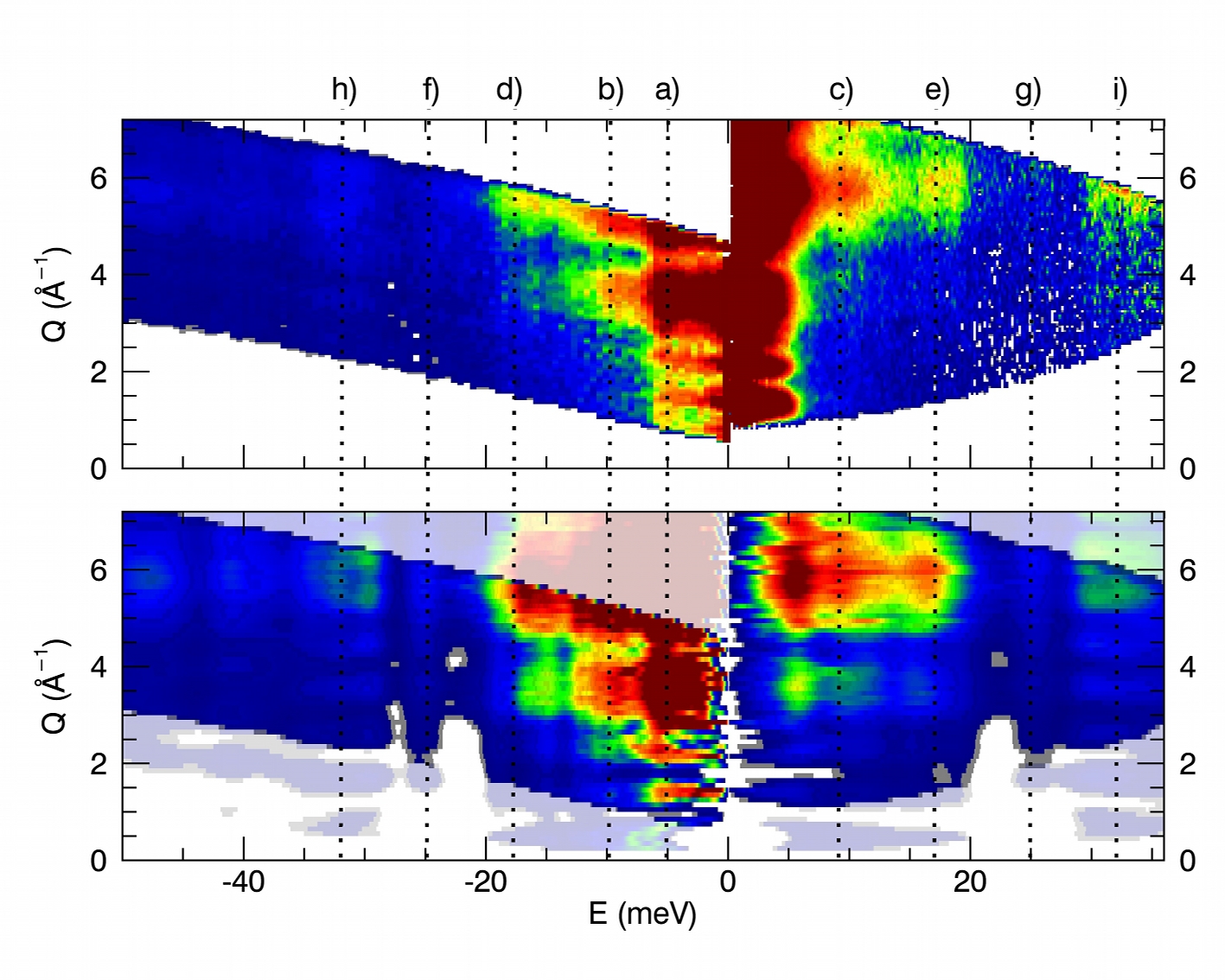}
\caption{$S(Q,\omega)$ (E = $\hbar\omega$) maps of Li$_4$C$_{60}$ measured at 330 K and 2.4 \AA\ (top left panel, with negative energy transfer), and at 10 K and 1.4 \AA\ (top right panel, with positive energy transfer). The corresponding simulated PALD maps are shown in the bottom panel for comparison. The spectra were normalized to the integrated intensity in the [16, 19 meV] regions.}
\label{fig:sqe_plots}
\end{figure}

In Fig.~\ref{fig:sqe_plots} the measured $S(Q,\omega)$ of Li$_4$C$_{60}$ collected at 330 K on IN4 and using an incident wavelength of 2.41 \AA\ is shown on the top left panel. It is compared to that obtained using a 1.4 \AA\ incident wavelength at 10 K (right panel). The bottom panel of Fig.~\ref{fig:sqe_plots} corresponds to the PALD simulated $S(Q,\omega)$ in the same conditions (E = $\hbar\omega$). 
The main characteristic features of the data are very well reproduced by the simulations, over a wide (Q, $\omega$) range, both in term of E and Q dependence of the spectra.
In particular, one observes a strongly Q-modulated signal, typical of coherent scattering, with maxima of intensity located around 3.4 and 5.7 \AA$^{-1}$. These features are commonly observed in fullerene systems (see \textit{e.g.} Ref.~\onlinecite{rolsepjst} and reference therein) reflecting the coherent scattering of carbon and high symmetry of the fullerene cages. They are typical of modes involving rotations of the cages. At low Q and $\omega$, one observes an intense signal emerging form the elastic scattering at $\sim$ 0.75, 1 and 2.2 \AA$^{-1}$ in the experimental spectra. These features correspond to acoustic branches which extend up to $\sim$ 6 meV according to our simulations. It is interesting to observe that a Q--cut of the intensity at an energy of 6 meV shows a rotation-like Q dependence with maxima at 3.4 and 5.7 \AA$^{-1}$. This suggests a strong hybridization between modes of translational and rotational character at the Brillouin zone edges, \textit{e.g.} strong rotation-translation coupling of the cages. This is also observed for modes in the intermediate energy range ($\hbar\omega \lesssim$ 19 meV) where both maxima are observed, while the higher frequency modes ($\hbar\omega \geq$ 30 meV) have Q dependences with a completely different scheme. 

\begin{figure}
\includegraphics[width=1.\textwidth]{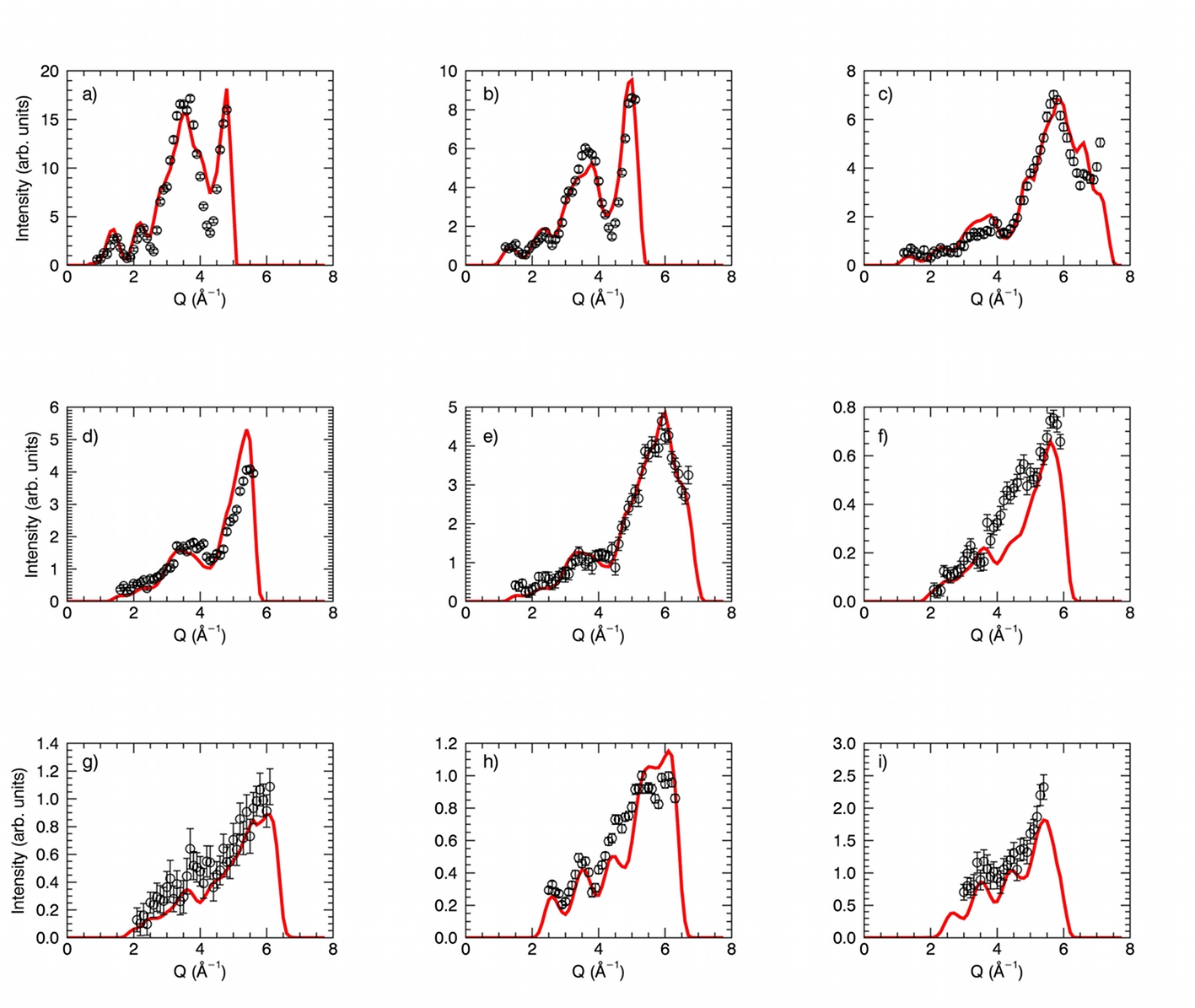}
\caption{Energy cuts (E = $\hbar\omega$) along selected lines in the S(Q,$\omega$) map according to Fig.~ \ref{fig:sqe_plots}: a) E = -5 meV, dE = 2 meV; b) E = -9.5 meV, dE = 4 meV; c) E = 9.5 meV, dE = 4 meV; d) E = -17 meV, dE = 4 meV; e) E = 17 meV, dE = 4 meV; f) E = -25 meV, dE = 6 meV; g) E = 25 meV, dE = 6 meV; h) E = -32 meV, dE = 8 meV; i) E = 32 meV, dE = 8 meV}
\label{fig:cuts_plots}
\end{figure}

\subsubsection{GDOS G($\omega$) in the polymer state}

\begin{figure}
\includegraphics[width=1.\textwidth]{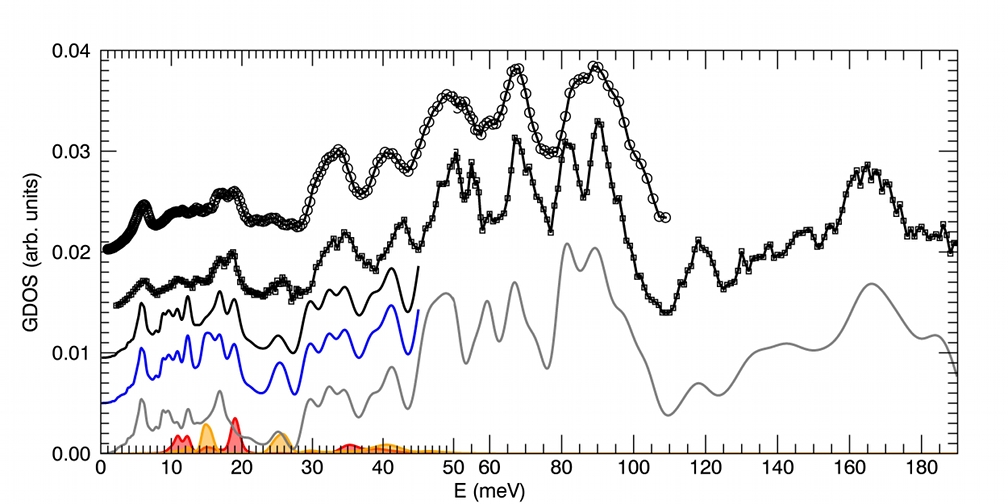}
\caption{Generalised density of states derived from inelastic neutron scattering data collected at 320 K (incident neutron wavelength 2.4 \AA\ - circles top curve) and 10 K (IN4C for energy lower than 60 meV: incident neutrons 2.2 \AA, 1.4 \AA\ and 1.1 \AA. IN1BeF for energy higher than 60 meV - squares). Data are compared to calculations (solid lines from top to bottom): GDOS of carbon and Li$_T$ (black line) and total GDOS (blue line). Partial GDOS for C (grey line), Li$_T$ (red line and area) and Li$_O$ ( yellow line and area) are also shown. For clarity, the GDOS were shifted by $\sim$ 0.005 units, except for the partial GDOS. For the same reason, the calculated spectra are not shown for frequency larger than 45 meV, the corresponding  spectra in that range being totally dominated by the C partial GDOS.}
\label{fig:GDOS_exp_sim}
\end{figure}

The Generalised Density of States (GDOS) of Li$_4$C$_{60}$ was extracted from $S(Q,\omega)$ following the standard data treatment described elsewhere \cite{Wdowik10}, using the ``incoherent approximation'' - it is shown in Fig.~\ref{fig:GDOS_exp_sim}.

Two datasets are shown. One was measured at 320 K using a relatively large incident wavelength of 2.4 \AA\ in anti-Stokes mode. The other was measured at 10 K in Stokes mode, which is composed of several datasets, measured on IN4C and IN1BeF spectrometers, in order to cover the whole spectrum. To match the partial datasets, they were normalized in common energy ranges.
The 320 K and 10 K spectra show the same characteristics: a first feature at 7 meV, followed by a range dominated by a double peak structure at 17 and 19 meV. In the pseudo-gap region, a feature is observed at 24 meV which seems better defined at low temperature, while significantly softened and split at 320 K. This mode was observed in the HREELS spectrum of a thin film of Li$_4$C$_{60}$ \cite{macovez2008}. At higher frequencies, the GDOS features the typical fulleride molecular modes.\\
The data are compared to the GDOS calculated for the model developed in the previous section. Two spectra are shown (blue and black lines respectively): the first represents the total GDOS, \textit{i.e.} comprising the sum of the C (grey line), Li$_T$ (red line and area) and  Li$_O$ ( yellow  line and area) partial GDOS contributions. The second one represents only the sum of the C and Li$_T$ contributions, which would correspond to the case in which the Li$_O$ atom vibrations were absent or distributed smoothly over a large spectrum range, \textit{i.e.} reflecting structural disorder.

The calculated spectra are in very good agreement with the data over the whole frequency range. In particular, the principal low frequency features of the experimental data are reproduced. The first peak of the doublet (at 17 meV) can be attributed to C$_{60}$ vibrations, while the second peak of the doublet matches a significant feature of the Li$_T$ partial GDOS. The feature at 24 meV can be attributed to a hybrid mode, with carbon and Li$_O$ contributions. However, the corresponding contribution in the total GDOS is significantly more intense than what is observed in the data. The double peak feature is also much less defined in this spectrum. A much better agreement with the experimental GDOS is obtained for the black spectrum: the 24 meV peak is less intense (with regards \textit{e.g.} to the intensity at 33 meV) and comparable to what is observed, and the double peak feature is much better defined. This suggests that the Li$_O$ contribution to the GDOS is not really observed in the data, reflecting the disordered nature of the Li$_O$ sub-lattice, in good agreement with the diffraction data. Polarized neutron investigations would be useful to isolate the smooth, broad distribution of Li modes as they allow the coherent (mostly C) and incoherent (essentially Li) scattering to be separated.\\


\begin{figure}
\includegraphics[width=0.85\textwidth]{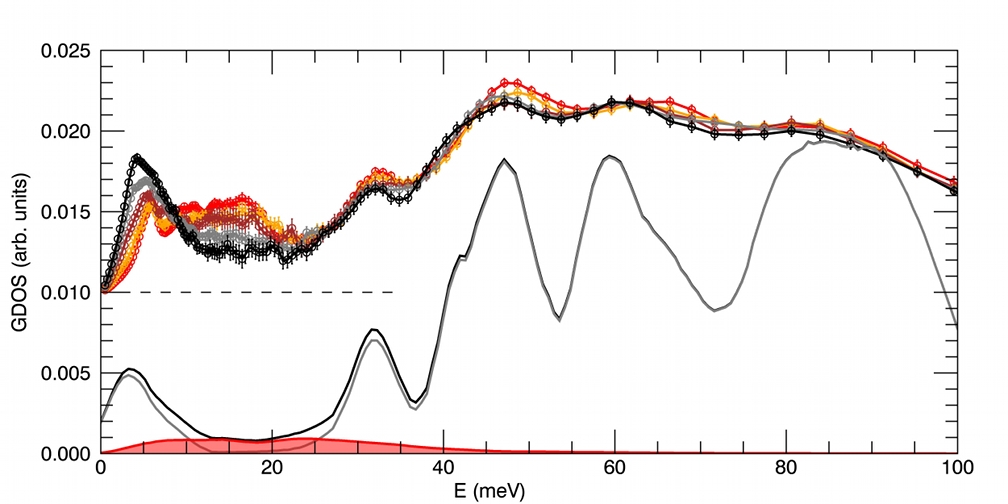}
\caption{Top: Generalised density of states (GDOS) derived from inelastic neutron scattering data collected between 300 K (polymer phase) and  700 K (monomer phase): red = 300 K, yellow = 610 K, brown = 630 K, grey = 640 K and black = 700 K). Bottom: GDOS extracted from the MD simulations at 800 K (see text) in the monomer phase (black solid lines: total GDOS, grey solid line: carbon GDOS, red area Lithium GDOS).}
\label{fig:GDOS_exp_mib}
\end{figure}

\subsubsection{Transformation to the monomer phase}

After the Li$_4$C$_{60}$ sample was heated progressively to 700 K, the polymer phase was transformed into a monomeric phase. Figure \ref{fig:GDOS_exp_mib} shows the evolution of the GDOS with temperature, obtained on the Mibemol spectrometer using a 5 \AA\ incident wavelength. It is compared to the total and partial GDOS derived from MD simulations. The spectral weight is observed to be transferred to low frequency in the monomer phase compared to the polymer phase, in agreement with a change in the intermolecular force constant. In particular, the modes in the [10, 30] meV disappear. The simulations show a very good agreement with the experimental spectra, in particular concerning the frequency and shape of the first intramolecular peak at 33 meV. The GDOS in the gap region (\textit{i.e.} in the [10, 30] meV range) is however not nul. According to the simulation, it contains contributions from the Li atoms, albeit more structured in the data than in the simulations. Phase segregation into Li and C$_{60}$ monomer clusters is probably responsible of this excess structured features in the gap.

\section{Conclusions}
\label{sec:conclusions}
In this paper we have reported an extensive neutron scattering investigation of the Li$_4$C$_{60}$ compound. Diffraction data defines a structure with a certain degree of disorder associated with one kind of Li atom, referred to as Li$_O$.\\
The lattice dynamics simulations give phonon dispersion curves and phonon DOS and identify Raman and IR active modes. The PALD method allows the $S(Q, \omega)$ maps of the system and the Generalysed Density of States to be generated and compared with neutron scattering observables - the agreement is very good.
The DFT simulations are coherent with a 3e charge transfer from the Li to the C$_{60}$ molecule and the peculiar polymeric bonding scheme proposed initially from diffraction and Raman investigations. The inelastic features from the Li$_O$ atoms are not observed in the experimental data, suggesting a disordered Li$_O$ sublattice. The presence of hybrid Li--C modes at relatively low frequencies, associating Li translations and either C$_{60}$ rotations or cage deformations, are highlighted. These modes have to be considered as part of the atomic description of the ionic diffusivity in this solid. In particular, a peak in the GDOS at 24 meV, attributed to an hybrid C and Li$_O$ feature, is found to be temperature dependent. The detailed temperature dependence of this peak should be used as a probe of the influence of the Li disorder and large amplitude C$_{60}$ modes on the ionic conduction in this system as it could reveal the microscopic origin of the Li mobility of this crystal.

\begin{acknowledgments}
The authors acknowledge the Institut Laue Langevin and the Laboratoire L\'{e}on Brillouin for neutron beam time allocation. We thank A. Ivanov and O. Meulien for scientific and technical assistance during the experiments on the IN1BeF and IN4C spectrometers. The DFT-VASP simulations were performed using the ILL licence allocated to the Computing for Science group at the ILL. Financial support from NMI3 is acknowledged.%
\end{acknowledgments}

\end{document}